\documentclass[aps,prl,twocolumn,showpacs]{revtex4}
\usepackage{graphicx,amsmath,amssymb}
\usepackage{hyperref}
\usepackage{bm}
\usepackage{color}

\begin{document}

\title{Unconventional  Filling Factor 4/11: A Closed-Form Ground State Wave Function  }	
\author{Sahana Das$^1$, Sudipto Das$^2$,  and Sudhansu S. Mandal$^{1,2}$}
\affiliation{$^1$Centre for Theoretical Studies, Indian Institute of Technology, Kharagpur 721302, West Bengal, India}
\affiliation{$^2$Department of Physics, Indian Institute of Technology, Kharagpur 721302, West Bengal, India}
\date{\today}
\begin{abstract}
 The ground state at 4/11 filling factor is very well understood [Phys. Rev. Lett. \textbf{112}, 016801 (2014)] in terms of the 1/3 filled second effective Landau level of the composite fermions whose correlations resemble with that of electrons in the ground state of two-body Haldane pseudo-potential of relative angular momentum 3, $V_3$. We here propose a closed-form ground state wave function for $V_3$ at 1/3 filling factor. We successfully compare it with the exact wave function for the systems with a few electrons, by calculating their mutual overlap, pair-correlation function, and entanglement spectra. By numerical exact diagonalization for a few electron systems, we find a window of nonzero $V_3$ is essential together with $V_1$  for being 4/11 state incompressible. The constructed wave function for 4/11 state using this proposed wave function has satisfactorily high overlap with the previously studied composite-fermion-diagonalized ground state wave function.      
\end{abstract}	
\maketitle

\section{Introduction}

Most of the fractional quantum Hall effect (FQHE) \cite{fqhe,laugh83} in the lowest Landau level (LL) belonging to the sequences of filling factors $\nu = n/(2pn\pm 1)$ and $1-n/(2pn\pm 1)$ are generally understood as $\nu^\ast= n$ integer quantum Hall effect \cite{iqhe} (integer number of filled effective LL called $\Lambda$ level) of composite fermions (CFs) \cite{jain89,jain_book} carrying $2p$ vortices denoted as $^{2p}$CFs. Amongst many other unconventional FQHE states in the lowest and higher LLs, the FQHE states in the range $ 1/3 < \nu <2/5$ are particularly intriguing \cite{Csathy20}. The states such as  $4/11$, $5/13$, $3/8$ and $6/17$ within this range are observed in the experiments \cite{Pan03,Pan15,Csathy15}, although latter two are not yet fully confirmed as there are no hint of flattening of the corresponding Hall resistances. These states correspond to respective fractional fillings $\nu^*=1+1/3$, $1+2/3$, $1+1/2$, and $1+1/5$ of $^2$CFs, i.e., $\Lambda=0$ is completely filled and $\Lambda =1 $ is partially filled with respective fractions $\bar{\nu}=1/3$, $2/3$, $1/2$, and $1/5$.
The neutral modes of excitations of these states display extremely low magneto-roton energies.\cite{Mandal2015} 
 While $\bar{\nu}=1/5$ seems  to be a conventional \cite{Balram} FQHE of CFs, the correlations for other three states can only be understood through unconventional \cite{sutirtha38,sutirtha411,sutirtha411b} mechanisms: (i) Moore-Read Pfaffian \cite{MR} correlation which is the exact ground state for a short-ranged three-body potential at $\bar{\nu}=1/2$ , (ii) Wojs-Yi-Quinn (WYQ) correlation \cite{wyq}, i.e., the ground state of two-body Haldane pseudo-potential \cite{haldane} $V_3$, (where $V_m$ is the model potential with two-body relative angular momentum $m$), at $\bar{\nu}=1/3$ and its particle-hole conjugate partner $2/3$. However, the absence of a suitable trial wave function of $\bar{\nu} = 1/3$  in the literature eludes us for knowing a closed-form ground sate wave function for $\nu =4/11$ and further investigations of its properties.

Our primary focus in this paper is proposing a trial wave function of $1/3$ WYQ state through several validity checks such as overlap with the exact wave function for a few electron systems, pair-correlation function \cite{Girvin1984} and qualitative low-energy features of the corresponding neutral mode in the single-mode approximation \cite{GMP1986}, and entanglement spectra \cite{Li2008}. Unlike the Laughlin state \cite{laugh83}, the counting of states in the entanglement spectra is found to be consistent with two Abelian edge modes \cite{Wen}. The satisfactorily well trial wave function for WYQ state enables us to construct a closed-form ground state wave function at $\nu= 4/11$ that has high overlap with the previously found composite-fermion diagonalized (CFD) ground state \cite{sutirtha411}. In addition, we investigate why wave function for $V_3$ model potential is necessary for understanding $4/11$ state while its neighboring conventional states $1/3$ and $2/5$ are well understood \cite{jain_book} only through the model potential $V_1$.


A two-body interaction operator $\hat{V}$ for fermions confined in the lowest LL, in general, may be expressed in  terms of two-particle projection operators $\vert m\rangle \langle m \vert$ as $
\hat{V} = \sum_{m\,({\rm odd})} V_m \vert m\rangle \langle m \vert $, 
where $\vert m\rangle$ denotes the two-particle state with relative angular momentum $m$, and $V_m$ is the so-called Haldane pseudo-potential \cite{haldane} describing energy of the state $\vert m \rangle$. For the Coulomb interaction in the lowest LL, $V_1$ dominates over other pseudo-potentials. The Laughlin wave  function at $\nu = 1/3$ with flux-shift $3$, i.e., number of flux quanta, $N_\phi = 3N-3$, is the exact ground state of $V_1$ and the other conventional FQHE states with filling factors $\nu = n/(2n\pm 1)$ can be reproduced by the model potential $V_1$ only. The Laughlin wave function for $\nu = 1/5$ with $N_\phi = 5N-5$ is the zero-energy ground state for $V_1 = V_3\neq 0$ and vanishing other higher order pseudo-potential components. The WYQ state\cite{wyq} at $\nu = 1/3$  corresponding to the ground state of $V_3$ occurs for $N_\phi = 3N-7$. The pseudo-potential $V_3$ for CFs  dominates over $V_1$ for the effective interaction between $^2$CFs in the second $\Lambda$ level. \cite{Sitko96,Scarola2001}
Therefore, the FQHE of $^2$CFs in $\Lambda=1$ should primarily be feasible for $V_3$ only. This is why the ground state wave function for $\nu =4/11$ that corresponds to $1/3$ FQHE of $^2$CFs in $\Lambda=1$ are well-described \cite{sutirtha411} by the WYQ correlation. 

 In section II, we begin with ruling out a simple possibility of trial wave function like an extension to  
 the CF wave function when only the third effective Landau level is completely filled, as trial wave function for the WYQ state at $\nu =1/3$. We then propose a successful trial wave function for this state as we find its reasonably high overlap with the exact ground state up to 13 electrons. We have also shown that overlap with the exact wave function may further be substantially improved by incorporating simple extension of this proposed wave function by their suitable superpositions. For further checking of the consistency of the proposed wave function, we calculate pair-correlation function, neutral mode of excitation within the single-mode approximation, and entanglement spectra that are qualitatively and even quantitatively close to that for the exact state. Although minimum gap for the neutral mode is an order of magnitude lower than the Laughlin state at the same filling factor, the finite gap ensures that the wave function represents an incompressible state.
 The low-lying entanglement spectra indicates the label counting of the edge states as $1,2,5,\cdots$ suggesting two abelian edge modes \cite{Wen} for the WYQ 1/3 state. The trial wave function for WYQ 1/3 state is then used to construct a trial wave function for 4/11 state in section III. This wave function has been shown to have high overlap with the CFD \cite{ssm1} ground state \cite{sutirtha411} which is close to the exact state. In section IV, we obtain a phase diagram in $V_1$--$V_3$ parameter space and identify the region for which $4/11$ becomes an incompressible state. The phase diagram indicates that an window of $V_3$ needs to be essentially mixed with $V_1$ for an incompressible 4/11 state, in consistence with our constructed wave function which consists of a part that is incompressible for $V_3$ pseudo-potential. Section V is devoted for a discussion about future direction of study. In  appendix A, we have developed a method how a manybody wave function in spherical geometry can be decomposed into the linear combination of determinants of occupied single particle basis states. In appendix B, we have reexpressed single particle basis functions \cite{jain_book} of the lowest two $\Lambda$ levels in a form \cite{Mandal2018} which shows similarity with the basis functions in a disc geometry \cite{jain_book}. Appendix C shows how a manybody wave function for the lowest Landau level can be recast for $\Lambda =1$ level.

\section{Trial Wave function for WYQ 1/3 State}

\begin{table}[h]
	\caption{Overlaps of the wave function $\Psi^{1/3}_{V_3}$ with $\Psi^{1/3}_{\rm CF,2}$, $\Psi^{1/3}_{\rm L-R}$ and $\Psi^{1/3}_{\rm L-MR}$	including appropriate normalizations at $\nu = 1/3$ for $N$ electrons.The numbers in $(..)$ indicate the Monte Carlo uncertainty in the last significant digits. $^\ast$For N= 5, $\Psi^{1/3}_{\rm CF,2}$  identically vanishes. $\Psi^{1/3}_{\rm L-R}$ is exact for $N=5$.} 
	\begin{tabular}{cccc}\hline\hline
		N \,\,& \,\,$\langle \Psi^{1/3}_{V_3}\vert \Psi^{1/3}_{\rm CF,2}\rangle$ & \,\,$\langle \Psi^{1/3}_{V_3}\vert \Psi^{1/3}_{\rm L-R}\rangle$ & \,\,$\langle  \Psi^{1/3}_{ V_3}\vert \Psi^{1/3}_{\rm L-MR}\rangle$ \\ \hline
		$5^\ast$ & -	& 1.0 & -\\ 
		6 & 0.299(2) & 0.89017 & 0.99762 \\
		7 & 0.213(2) & 0.91661 & 0.95505 \\
		8 & 0.205(3) & 0.88519 & 0.92107 \\
		9 & 0.158(5) & 0.76491 & 0.90428 \\
		10 & - & 0.72486 & 0.86071\\
		11 & - & 0.753(4) & 0.863(2)\\
		12 & - & 0.773(6) & 0.854(2)\\
		13 & - & 0.781(3) &0.836(2) \\ 
		14 &-  & 0.777(2) & 0.824(1)\\ \hline
	\end{tabular}
	\label{Table1}
\end{table}

Since the WYQ 1/3 state occurs for $N_\phi = 3N-7$ in contrast to $N_\phi = 3N-3$ for Laughlin wave function, 
\begin{equation}
\Psi_L^{1/3}=\prod_{i<j}(u_iv_j-u_jv_i)^3
\label{Eq1}
\end{equation} 
which is also same as the CF wave function \cite{jain89,jain_book},
it is tempting to write a trial wave function 
\begin{equation}
\Psi_{\rm CF,2}^{1/3} = {\rm P_{LLL}}\,\prod_{i<j} (u_iv_j-u_jv_i)^2 \,\chi_2(\{u_i,\,v_i\})
\label{Eq3}
\end{equation}
 where $\chi_2$ is the wave function for $\Lambda=2$ level being completely filled by the CFs while keeping the lower $\Lambda$ levels with $\Lambda =0$ and $1$ completely empty, and ${\rm P_{LLL}}$ represents projection onto the lowest LL.
 Here $u_j = \cos (\theta_j/2) e^{i\phi_j/2}$ and $v_j = \sin (\theta_j/2) e^{-i\phi_j/2}$ are the spherical spinors in terms of spherical coordinates for $j^{\rm th}$ electron in a spherical geometry \cite{haldane} of radius $R = \sqrt{Q}$ in the unit of magnetic length, $\ell= (\hbar c /eB)^{1/2}$, with magnetic monopole charge $Q= N_\phi/2$ residing at the center of the sphere.
 Since the overlap of $\Psi_{\rm CF,2}$ with the exact ground state has not been found to be impressive (Table~\ref{Table1}), it cannot be considered as satisfactory trial wave function.

\begin{figure}
	\centering
	\includegraphics[scale=1.3]{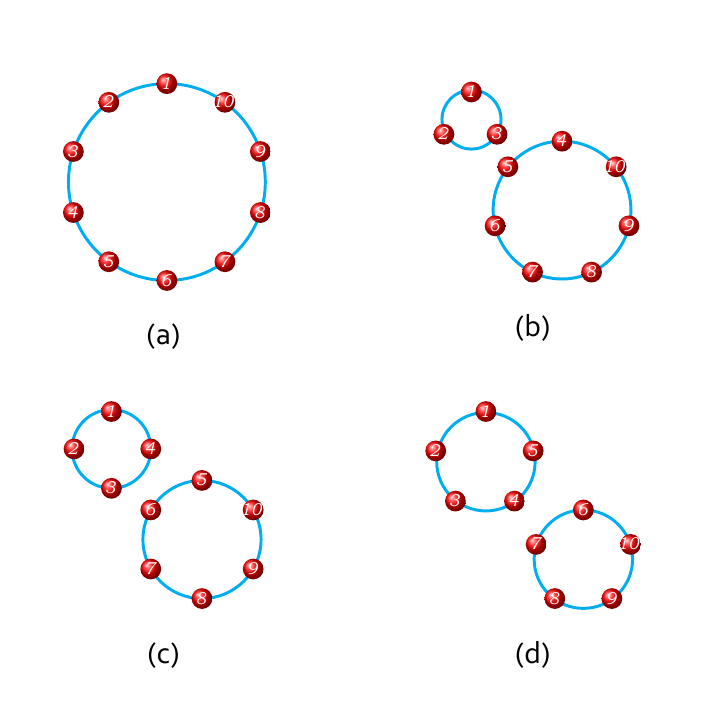}
	\caption{(Color online) Schematic arrangement of 10 particles in closed rings where spheres represent electrons and the connecting lines between $i^{\rm th}$ and $j^{\rm th}$ electrons represent the function $(u_iv_j-v_iu_j)^{-2}$. Diagrammatic representations of the ring functions: (a) ${\cal R}_{10}$, (b) ${\cal R}_7{\cal R}_3$, (c) ${\cal R}_6{\cal R}_4$, and (d) ${\cal R}_5{\cal R}_5$.}	
	\label{Fig1}
\end{figure}

We here propose that the ground state wave function of $V_3$ at $1/3$ as
\begin{eqnarray}
\Psi^{1/3}_{\rm  \small L-R} (\{u_i,\,v_i\})&=& \prod_{i<j}^N (u_iv_j-v_iu_j)^3\, {\cal S}\,\left({\cal R}_N\right) \, ; \label{Eq4}\\ {\cal R}_N  &=&\left[ \prod_{i=1}^N (u_iv_{i+1}-v_iu_{i+1})^{-2}\right]
\label{Eq5} 
\end{eqnarray}
where the extra factor ${\cal R}_N$ represents a ring-correlation between $N$ electrons; electrons are arranged in a closed ring (see Fig.~\ref{Fig1}) with $u_{_{N+1}}= u_{_1}$ and $v_{_{N+1}}=v_{_1}$.
Here ${\cal S}$ represents the symmetrization for $N$ identical particles. It is easy to check that the angular momentum of the wave function (\ref{Eq4}), $L=0$. Although it appears singularity in ${\cal R}_N$ when two electrons are closed, it is removed in $\Psi^{1/3}_{L-R}$ and Pauli exclusion principle is restored. The corresponding wave function in the disc geometry will read (by dropping ubiquitous Gaussian factor) as
\begin{equation}
\Psi_{\rm L-R}^{1/3}(\{z_j\}) = \prod_{i<j}^N (z_i-z_j)^3 \,{\cal S}\left(\prod_{i=1}^N (z_i-z_{i+1})^{-2}  \right)
\label{Eq6}
\end{equation}
with $z_j = (x_j -i y_j)/\ell$ and $z_{N+1}=z_1$.

\begin{table} 
	\caption{Weight factors with signs of different normalized ring wave functions (up to two rings) in $\Psi_{\rm L-MR}^{1/3}$. We note that the sum of the square of the weight factors are not necessarily one as the ring wave functions are not mutually orthogonal. }
	\begin{tabular}{|c|c|c|c|c|c|c|}\hline\hline
	$N$ & ${\cal R}_N$ & ${\cal R}_{N-3} {\cal R}_3$ & ${\cal R}_{N-4} {\cal R}_4$ & ${\cal R}_{N-5} {\cal R}_5$ & ${\cal R}_{N-6} {\cal R}_6$ &  ${\cal R}_{N-7} {\cal R}_7$ \\ \hline
	5 & 1.0 & --- &--- &--- &---&--- \\
	6 & 0.779 & $-0.465$ & --- &--- & ---&--- \\
	7 & 0.817 & $-0.315$ & --- &--- & ---&--- \\
	8 & 0.764 & $-0.315$ & $-0.126$ &--- & --- &---\\
	9 & 0.423 & $-0.612$ & $-0.196$ &--- & ---&--- \\
	10 & 0.374 & $-0.579$ & $-0.278$ &$-0.096$ & ---&--- \\
	11& 0.348 & $-0.473$ & $-0.372$ &$-0.160$ & ---& ---\\
	12& 0.380 & $-0.344$ & $-0.382$ &$-0.212$ & $-0.126$&--- \\
	13& 0.426 & $-0.231$ & $-0.313$ &$-0.242$ & $-0.175$& --- \\
	14& 0.414 & $-0.172$ & $-0.250$ &$-0.268$ & $-0.218$& $-0.147$ \\ \hline
	\end{tabular}
	\label{Table1a}	
\end{table}

\begin{figure}
	\centering
	\includegraphics[scale=0.65]{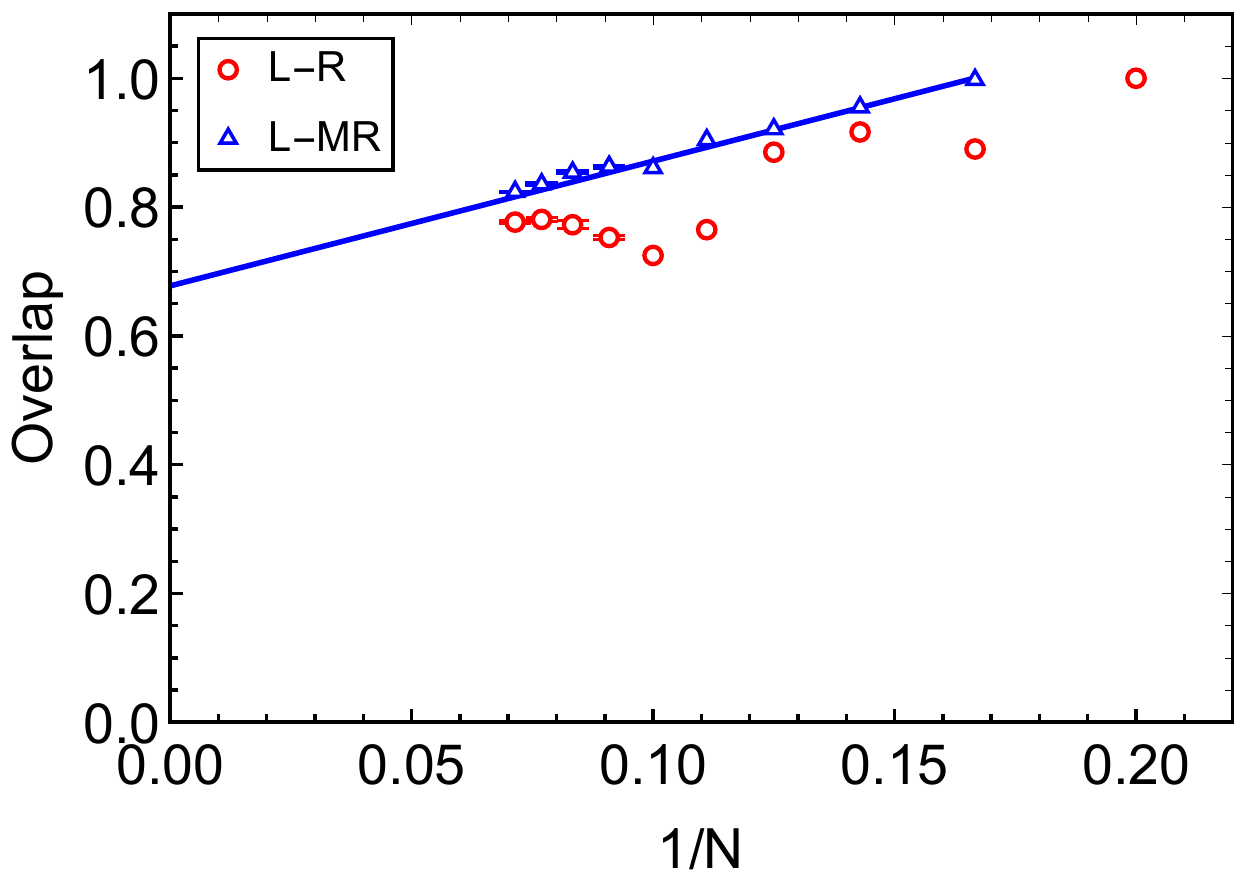}
	\caption{(Color online) Overlaps $\langle \Psi^{1/3}_{V_3}\vert \Psi^{1/3}_{\rm L-R}\rangle$  and $\langle  \Psi^{1/3}_{ V_3}\vert \Psi^{1/3}_{\rm L-MR}\rangle$ tabulated in Table~\ref{Table1} vs. $1/N$. The line is a guide to the eye. }	
	\label{Fig1a}
\end{figure}

We determine ground state wave function  $\Psi^{1/3}_{V_3}$  by exactly diagonalizing \cite{diagham} the pseudo-potential $V_3$ with $N_\phi = 3N-7$. The incompressible (the lowest energy state is at $L=0$ only) ground state is obtained for $N\geq 5$. The overlap between  $\Psi^{1/3}_{V_3}$ and $\Psi^{1/3}_{\rm L-R}$, $\langle \Psi^{1/3}_{V_3} \vert\Psi^{1/3}_{\rm L-R}\rangle$ obtained by the method of decomposition into single particle eigen basis (DSPEB) introduced first time here (see Appendix A) for smaller $N$ and by the Monte Carlo method in Metropolis algorithm for larger $N$ is tabulated in Table~\ref{Table1}. While the latter method has statistical uncertainty, the former provides exact value, albeit limited to lesser number of particles. We find that $\Psi^{1/3}_{L-R}$ is exact for $N=5$, but the overlap somewhat decreases with
the increase of $N$. However, much improved overlap is obtained by mixing functions ${\cal R}_N$ with ${\cal R}_{N-k}{\cal R}_k$ where $k=3,4,\cdots ,(N-1)/2$ $(N/2)$ for odd (even) $N$. Here $k_{\rm min}=3$ because ring is possible for at least three particles. These rings for $N=10$ are schematically shown in Fig.~\ref{Fig1} and the corresponding variational ground state wave function is denoted as $\Psi^{1/3}_{L-MR}$. 
The weight factors of the wave functions (not mutually orthogonal) constructed with ring functions ${\cal R}_{N-k}{\cal R}_k$ in $\Psi^{1/3}_{L-MR}$ are tabulated in Table~\ref{Table1a}.
As the construction of exact real space wave function in each Monte Carlo step becomes computationally expensive for larger $N$ due to exponential growth of basis states, we are able to compare our proposed wave function with the exact wave function up to $N=14$ only for which the number of basis states is $\sim 4.8\times 10^7$. The overlap 
$\langle \Psi^{1/3}_{V_3} \vert\Psi^{1/3}_{\rm L-MR}\rangle$  decreases with $N$, yet it seems to have reasonably high value in the thermodynamic limit (Fig.~\ref{Fig1a}). Although the overlap $\langle \Psi^{1/3}_{V_3} \vert\Psi^{1/3}_{\rm L-R}\rangle$ decreases very fast with the increase in $N$ up to $N=10$, but thereafter it slowly increases with $N$ and approaches  $\langle \Psi^{1/3}_{V_3} \vert\Psi^{1/3}_{\rm L-MR}\rangle$.
%
\begin{figure}[h]
	\vspace{0.5cm}
	\centering
	\includegraphics[scale=0.45]{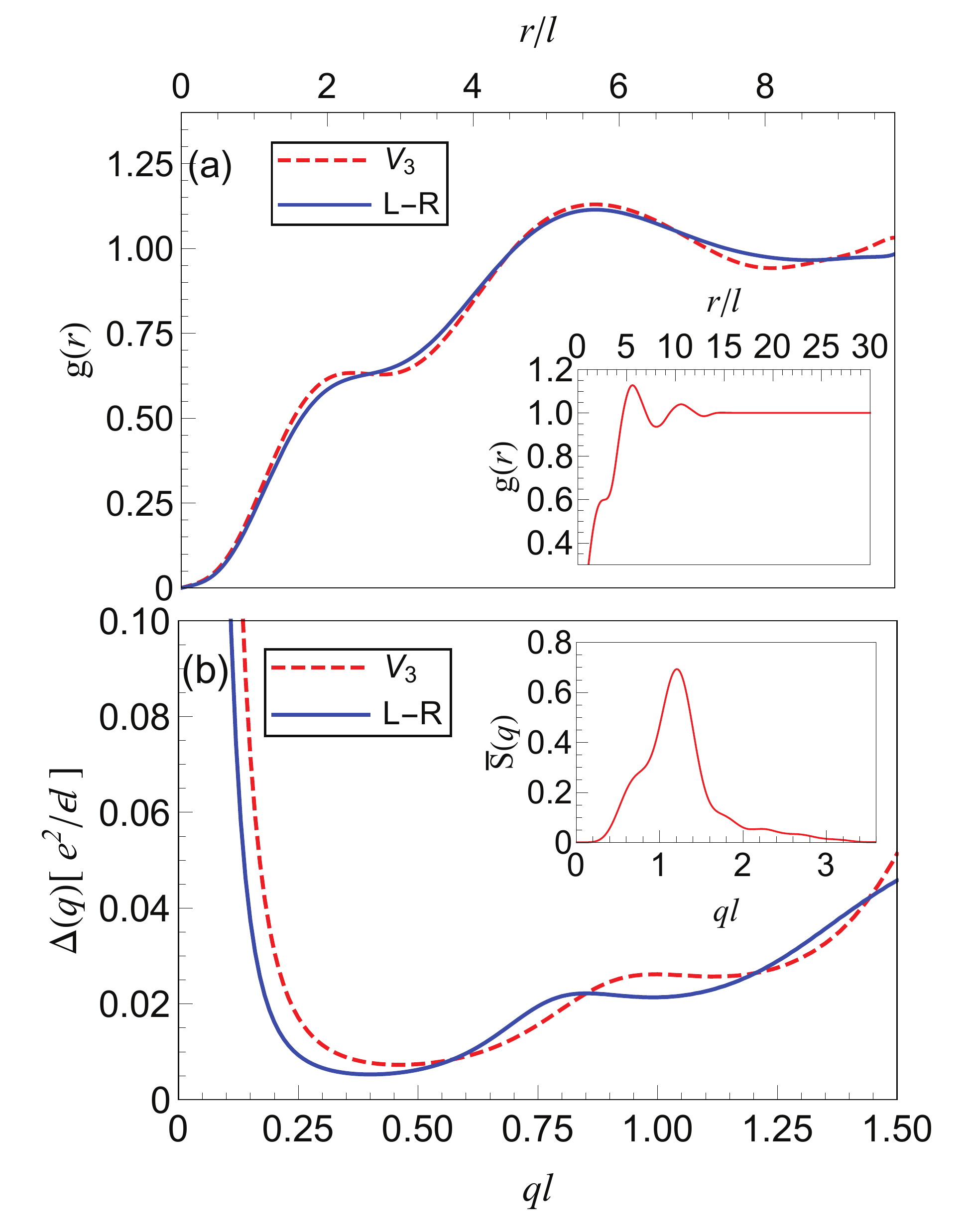}
	\caption{(Color online) Top panel: (a) Pair correlation function obtained for $N=10$ using the wave functions $\Psi^{1/3}_{V_3}$ (dashed line) and $\Psi^{1/3}_{L-R}$ (solid line); Inset: Thermodynamic extrapolation of $g(r)$ for the wave function $\Psi^{1/3}_{V_3}$ in the damped-oscillatory form\cite{kamilla1997} $g(r)=1+A(r/\ell)^{-\alpha} \sin(\beta r/\ell-\gamma)$ used earlier for the oscillatory part, where $A$, $\alpha$, $\beta$ and $\gamma$ are numerical constants. Bottom panel: (b) Dispersion of the GMP mode $\Delta (q)$; Inset: The lowest LL-projected structure factor $\bar{S}(q)$ calculated using thermodynamically extrapolated $g(r)$ and further fitting with the GMP form\cite{GMP1985,GMP1986} $g(r) = 1-e^{-r^2/(2\ell^2)} +\sum_{m\,(\rm odd)}(2/m!)(r^2/4\ell^2)^m c_m e^{-r^2/(4\ell^2)}$, where the coefficients $c_m$ (up to a suitable maximum value of $m$ for picking up the oscillations in $g(r)$) are to be determined by fitting.  }
	\label{Fig2}	
\end{figure}

\begin{figure}[h]
	\centering
\includegraphics[scale=0.41]{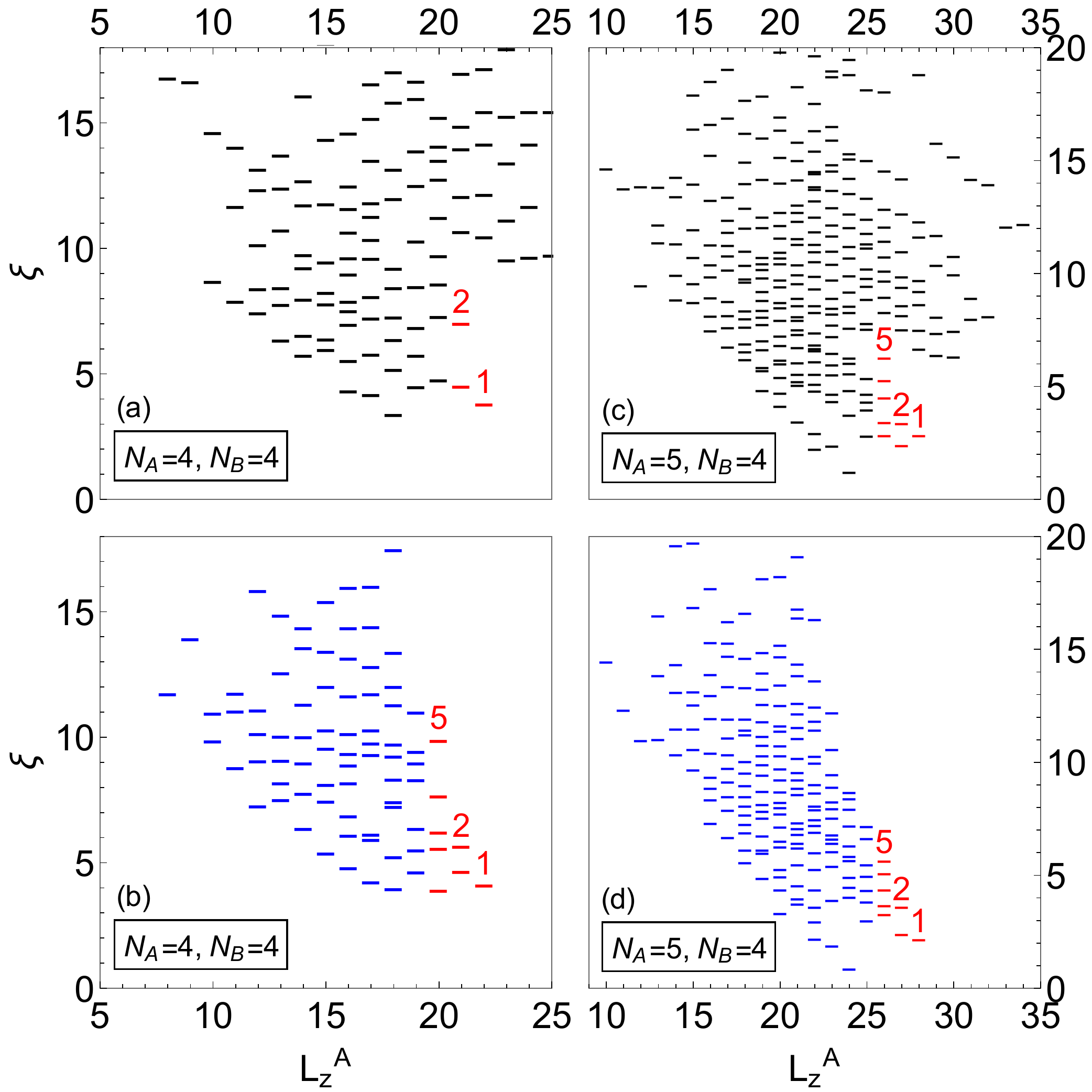}
	\caption{(Color online) Entanglement Spectra obtained by the method of particle partitioning in real space. $N_A \,(N_B)$ is the number of particles in northern (southern) hemisphere of the spherical geometry. $L_z^A$ is the $z$-component of the total orbital angular momentum of the subsystem A.
	(a) and (b) respectively for $\Psi^{1/3}_{V_3}$ and $\Psi^{1/3}_{\rm L-R}$ with $N_A=4$ and $N_B=4$. (c) and (d) respectively for $\Psi^{1/3}_{V_3}$ and $\Psi^{1/3}_{\rm L-R}$ with $N_A=5$ and $N_B=4$. The counting of low-lying levels (starting from the maximum value of $L_z^A=22$ for the system $N=8$ electrons and  $L_z^A=28$ for $N=9$) for $\Psi^{1/3}_{\rm L-R}$ is in agreement with that for $\Psi^{1/3}_{V_3}$.
	The low-lying levels are counted as 1, 2, 5,...
	}
	\label{Fig3}
\end{figure}

\begin{figure}[h]
	\centering
	\includegraphics[scale=0.66]{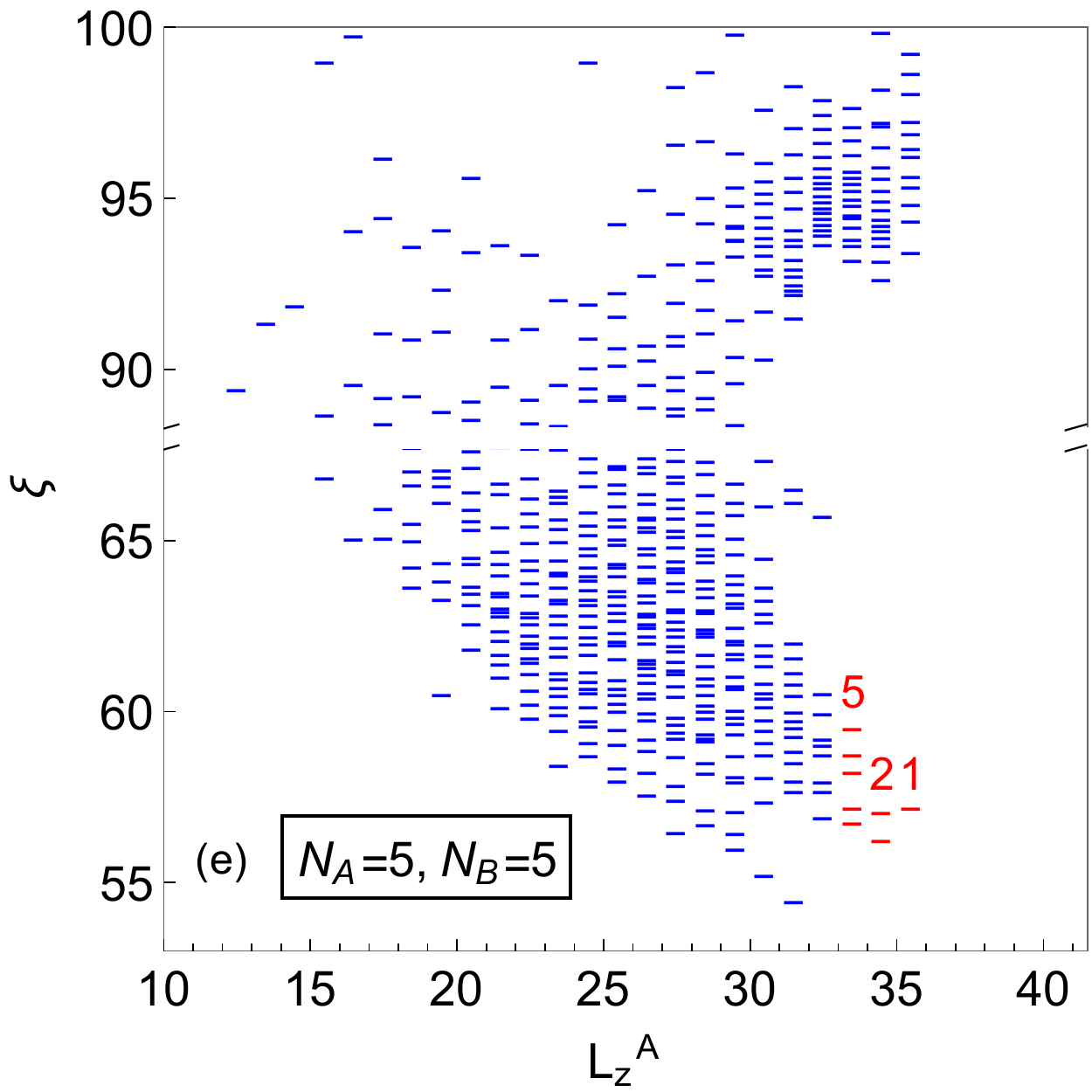}
	\caption{(Color online) Entanglement spectra with particle partitioning for $\Psi^{1/3}_{L-R}$ as described in the caption of Fig.~4. Here $N_A=N_B=5\, (N=10)$ and the low-lying spectra starts from maximum value of $L_z^A=35.5$. The low-lying level counting is found as $1, \,2, \,5,\, \cdots$, which matches with the counting of {\em two} chiral edge modes.\cite{Wen}}
	\label{Fig4}
\end{figure}

Having shown $\Psi^{1/3}_{\rm L-MR}$ is a variationally improved trial wave function above, we find below that $\Psi^{1/3}_{\rm L-R}$ is indeed a topologically sufficient trial wave function for WYQ 1/3 state by comparing the corresponding pair-correlation function, neutral mode of excitation, and the state counting in the low-lying sector of the entanglement spectra with that of the exact ground state. $\Psi^{1/3}_{\rm L-R}$ is thus adiabatically connected to $\Psi^{1/3}_{\rm L-MR}$ and $\Psi^{1/3}_{V_3}$, and in turn topologically distinct from $\Psi^{1/3}_{\rm L}$.

\subsection{Pair-Correlation and Neutral Mode}

In Fig.~\ref{Fig2}, we show pair correlation function $g(r) = \frac{1}{N\rho_0} \langle \sum_{i<j} \delta (\bm{r}-\bm{r}_{ij})  \rangle$ with mean electron density $\rho_0$ and inter-particle separation $\bm{r}_{ij}=\bm{r}_i -\bm{r}_j$ for $N=10$ in the ground states of $\Psi^{1/3}_{V_3}$ and  $\Psi^{1/3}_{\rm L-R}$. The simple trial wave function 
$\Psi^{1/3}_{\rm L-R}$ satisfactorily reproduces all the essential features of $g(r)$ that one finds with the exact state, in particular, the unusual (not present for Laughlin state\cite{Girvin1984}) hump appears (which usually found for nonabelian state like Moore-Read state \cite{ReadRezayi1996}) at $r \sim 2.5 \ell$. These $g(r)$ are then exploited to determine neutral modes of excitations by the method of single mode approximation which was originally developed by Girvin-MacDonald-Platzman (GMP)\cite{GMP1985,GMP1986}. The mode is determined by using previously derived expression\cite{GMP1985,GMP1986}
\begin{eqnarray}
	\Delta(q) &=& 2 \left(\bar{S}(q)\right)^{-1} \int \frac{d{\rm \bf k}}{(2\pi)^2} \text{sin}^2\left(\frac{{\rm \bf q}\times 
		{\rm \bf k}}{2}\ell^2 \right) e^{-q^2\ell^2/2} \nonumber \\
	&\times & \left[ v(\vert {\rm\bf k}-{\rm \bf q}\vert)e^{{\rm \bf q}\cdot    ({\rm \bf k}-{\rm \bf q}/2)\ell^2}-v(k) \right] \bar{S}(k)
	\label{Eq7}
\end{eqnarray}
where projected structure factor $\bar{S}(k)
= S(k) -1 +e^{-k^2\ell^2/2}$ and $v(q) = \left( \frac{2\pi e^2}{\epsilon q}\right) e^{-q^2\ell^2} L_3(q^2\ell^2)$ is the momentum-dependent potential \cite{Morf} corresponding to $V_3$ pseudo-potential component of the Coulomb interaction. Here $S(k) = 1 + n_0\int d\bm{r}e^{i\bm{k}\cdot \bm{r}} [g(r)-1]$ is the static structure factor, where the mean electron density $ n_0= \nu/ 2\pi \ell^2 $ and $L_3(q^2\ell^2)$ is the third order Laguerre polynomial.
The neutral modes for $\Psi^{1/3}_{\rm L-R}$ agrees quite well with that for $\Psi^{1/3}_{V_3}$. Unlike \cite{GMP1985} Laughlin wave function, both of these show two side-by-side roton minima and the minimum gap is much lower than the Laughlin state.

\subsection{Entanglement Spectra}

 The state counting in the low-lying entanglement spectra (ES)\cite{Li2008,Chandran2011,Rodriguez2012,Dubail2012,Sterdyniak2012,Rodriguez2013} has now been routinely used for determining the number of states at the edges \cite{Wen} of the FQHE systems.
 It therefore has been very useful for determining topological nature of a FQHE state. The entanglement spectrum of an incompressible ground state is characterized by an entanglement gap separating low-lying spectrum from the high-energy sector \cite{Thomale2010}.

 The ES are generally obtained by partitioning the system into two sub-systems in a number of ways, namely, orbital partitions, particle partition, and partition in real space. Here, we employ the method described in Ref.~\onlinecite{Rodriguez2012} by dividing the sphere into two hemispheres \textit{A} (upper hemisphere) and \textit{B} (lower hemisphere), so that the Fock space of the Hamiltonian $\mathcal{H}$ is partitioned into two parts $\mathcal{H}_A \otimes \mathcal{H}_B$. Using Schmidt decomposition \cite{Nielsen2000}, a many-body  ground state wave function for whole system can be decomposed into the linear combination of the products of states in two subsystems:
 \begin{equation}
 |\psi\rangle = \sum_i e^{-(1/2)\xi_i} |\psi_A^i\rangle \otimes |\psi^i_B\rangle \, , \label{Eq8}
  \end{equation}
 where, $  |\psi_A^i\rangle \in \mathcal{H}_A, \quad |\psi^i_B\rangle \in \mathcal{H}_B,\quad  \langle\psi_A^i|\psi_A^i\rangle = \langle\psi_B^i|\psi_B^i\rangle = \delta_{ij}$ and $\xi_i$ represents entanglement energy for $i^{\rm th}$ state. Therefore, $\xi_i$ can be obtained by diagonalizing the reduced density matrix for a subsystem, say $A $, i.e., $\hat{\rho}_A$ which may be obtained by tracing over \textit{B} degrees of freedom of the full density matrix: $\hat{\rho}_A = \text{Tr}_B [\hat{\rho}]$.
 If the subsystems contain $N_A$ and $N_B$ numbers of electrons respectively, the total azimuthal angular momentum of the subsystems, $L_z^{A/B} = \sum_{k=1}^{N_A/N_B} l_z^k$ with $l_z^k = -Q,\cdots ,+Q$ ($l_z^k$ is positive (negative) in $A\,(B)$). We determine entanglement spectra for each $L_z^A$ separately. In Fig.~\ref{Fig3}, we compare ES for $N=8$ and 9 electrons calculated using the trial wave function $\Psi^{1/3}_{\rm L-R}$ with that for the exact wave function $\Psi^{1/3}_{V_3}$ of WYQ state at 1/3 filling. The state-counting for the low-lying states in the spectra of $\Psi^{1/3}_{\rm L-R}$ matches that with  $\Psi^{1/3}_{V_3}$. The counting goes as $1,2,5,\cdots$ which is further confirmed (Fig.~\ref{Fig4}) in the spectra of N=10. This sequence of counting resembles with two abelian edge modes \cite{Wen}. This indicates that the WYQ 1/3 state has two edge modes rather than one as evidenced for the Laughlin 1/3 state. Two magneto-roton minima in the neutral mode (Fig.~\ref{Fig2}) for the bulk excitations is in consistent with the two edge modes.

\section{Closed-form Wave function for 4/11 state}

In Ref.\onlinecite{sutirtha411}, the ground state wave function for $\nu = 4/11$ was proposed as CF-WYQ wave function:
\begin{equation}
\Psi_{\rm CF-WYQ}^{4/11} = {\rm P}_{\rm LLL} \prod_{i<j}^{[1,N]} (u_iv_j-u_jv_i)^2 \,   \Phi^{1+1/3}_{\rm WYQ} \label{Eq9}
\end{equation}
where $\Phi^{1+1/3}_{\rm WYQ}$ is the determinant with $\Lambda = 0$ completely filled by $N-N^\ast$ particles and $1/3$-filled $\Lambda=1$ level with $N^\ast = N/4+2$. Here WYQ wave function for $1/3$ state is the exact numerical wave function (linear combination of $N^\ast \times N^\ast$ determinants when $N^\ast$ particles occupy certain single particle states) for the ground state of the pseudo-potential $V_3$.

As we now have a trial wave function (Eqs.~\ref{Eq4} and \ref{Eq5}) for  $\Phi^{1/3}_{V_3}$, we  explicitly construct the corresponding wave function for $\nu = 4/11$ as
\begin{eqnarray}
\Psi^{4/11} &=&  \prod_{i<j}^{[1,N]} (u_{i}v_j-u_{j}v_i)^2\, {\cal A} \left[ \prod_{k<l}^{[N^\ast +1,N]} (u_{k}v_l-u_{l}v_k)  \right. \nonumber \\
&\times& \left.\left(\prod_j^{[1,N^\ast]} Q_j \right)  \left( D_\alpha
\Psi^{1/3,N^\ast}_{\alpha,\rm L-R} (\{u_i,\,v_i\}) \right) \right] \label{Eq10}
\end{eqnarray}
Here anti-symmetrization ${\cal A}$ may be performed conveniently by multiplying the corresponding factor $(-1)^{\sum_{j} j}$ with each of the combinations $^NC_{N^\ast}$, where $j$ represents the particle number associated in the second $\Lambda$ level,
 the second $\Lambda$ level projection factor $Q_{j}  =\sum_{l\neq j}^{[1,N]} v_jv_l(u_jv_l-u_lv_j)^{-1}$ into the lowest $\Lambda$ level, and $D_\alpha$ represents numerical factor (see Appendix B and C for details) associated with $\alpha$ basis of the DSPEB of $ \Psi_{\rm L-R}^{1/3, N^\ast}$. However, this detailed numerical factors and needful of DSPEB is special for the spherical geometry. The proposed wave function in the disc-geometry will have much simpler structure:
 \begin{eqnarray}
 \Psi^{4/11}_{\rm L-R} &=& \prod_{i<j}^{[1,N]} (z_i-z_j)^2 {\cal A} \left[ \prod_{k<l}^{[N^\ast +1,N]} (z_k-z_l)\left( \prod_j^{[1,N^\ast]} P_j \right) \right. \nonumber \\
&\times& \left.  \Psi_{{\rm L-R}}^{1/3,N^\ast} (\{z_j\})\right]\,\exp\left[-\frac{1}{4\ell^2}\sum_j \vert z_j\vert^2 \right]
\label{Eq11}
 \end{eqnarray} 
 with $P_j = \sum_{l\neq j}^{[1,N]} (z_l-z_j)^{-1}$ and the explicit form of $\Psi_{\rm L-R}^{1/3,N^\ast}$ is shown in Eq.(\ref{Eq6}) for $N^\ast$ particles.
 
 \begin{figure}
 	\centering
 	\includegraphics[scale=0.4]{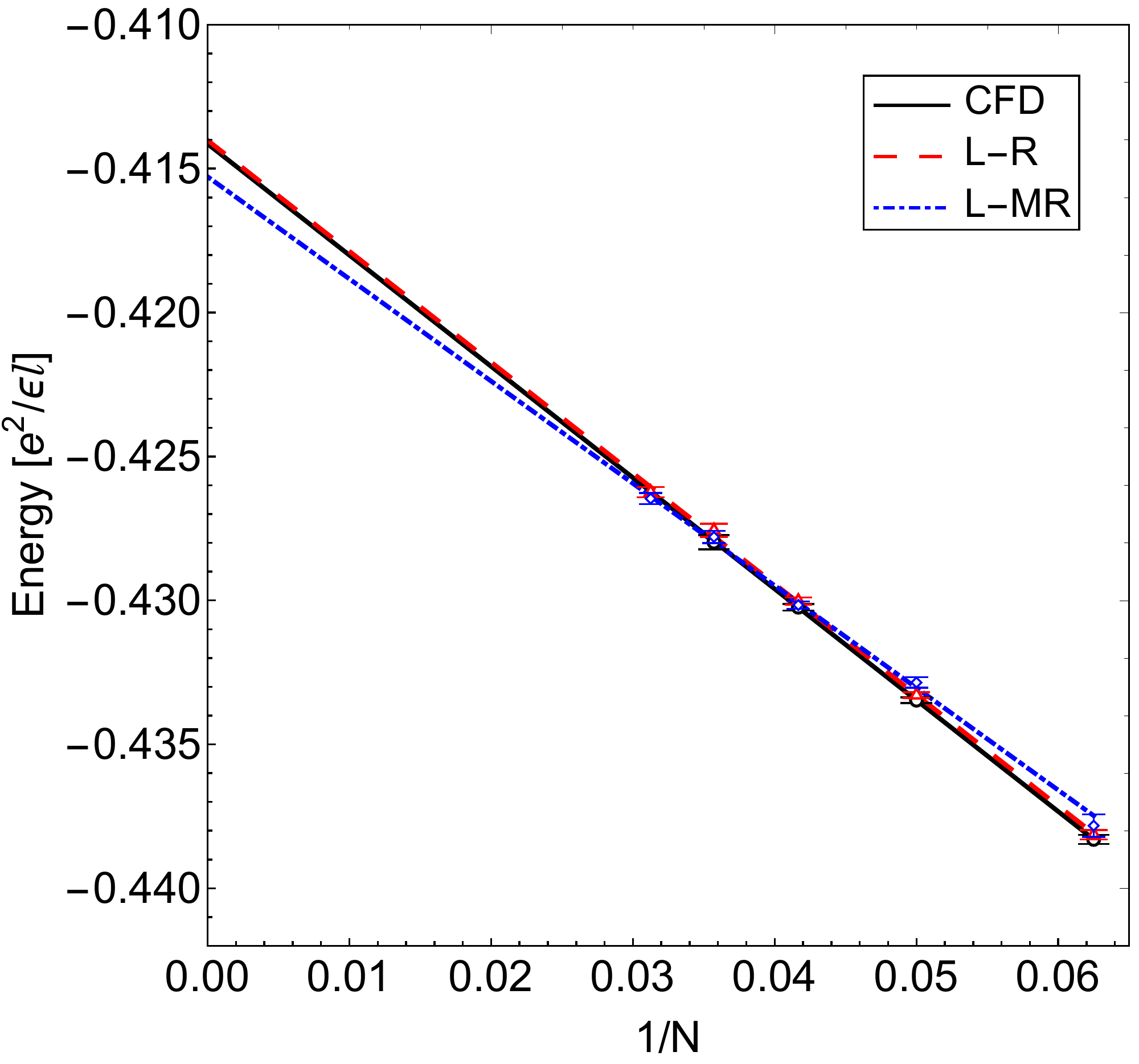}
 	\caption{(Color online) Ground state energy per electron versus $1/N$ for $\Psi^{4/11}_{\rm CFD}$ (circles), $\Psi^{4/11}_{\rm L-R}$ (triangles), and $\Psi^{4/11}_{\rm L-MR}$ (diamonds). The corresponding linearly fitted lines are extrapolated to determine the ground state energies in the thermodynamic limit.   }	
 	\label{Fig5a}
 \end{figure}
 
 The overlap of the trial wave function $\Psi^{4/11}_{\rm L-R}$ (\ref{Eq10}) with the  composite-fermion-diagonalized \cite{ssm1} wave function \cite{sutirtha411} $\Psi^{4/11}_{\rm CFD}$ for $\nu = 4/11$ is tabulated in Table~\ref{Table2}. The overlaps are reasonably high and it is further increased with the trial wave function $\Psi^{4/11}_{\rm L-R}$ replaced by $\Psi^{4/11}_{\rm L-MR}$. A comparison has also been made with the overlap $\langle \Psi^{4/11}_{\rm CFD} \vert  \Psi^{4/11}_{\rm CF-WYQ}\rangle$ reported earlier.\cite{sutirtha411}  
 In Fig.~\ref{Fig5a}, we compare the energies corresponding to $\Psi^{4/11}_{\rm CFD}$, $\Psi^{4/11}_{\rm L-R}$, and $\Psi^{4/11}_{\rm L-MR}$ states for various $N$. The linear extrapolations of these data determine the respective thermodynamic energies per particle as $-0.4141(4)$, $-0.4140(3)$, and $-0.4153(5)$ in the unit $e^2/(\epsilon \ell)$ where $\epsilon$ is the dielectric constant of the background. Surprisingly, the energy of the trial wave function is very close to that of the CFD wave function. As the overlaps 
 $\langle \Psi^{4/11}_{\rm CFD} \vert  \Psi^{4/11}_{\rm L-R}\rangle$ for the systems that we have studies are reasonably high and the ground state energies corresponding to these states in the thermodynamic limit are very close, the trial wave function Eqs.~(\ref{Eq10})  may be regarded as a good trial wave function  in spherical  geometry for $\nu =4/11$ state.
 
 \begin{table}
 	\caption{Overlaps of the wave function $\Psi^{4/11}_{\rm CFD}$ with $\Psi^{4/11}_{\rm CF-WYQ}$, $\Psi^{4/11}_{\rm L-R}$, and $\Psi^{4/11}_{\rm L-MR}$ for $N$ electrons of which $N^\ast$ electrons in $\Lambda =1$ level. The numbers in $(..)$ indicate the Monte Carlo uncertainty in the last significant digits.  }	
 	\begin{tabular}{ccccc}\hline\hline
 		$N\,\,$ & $N^\ast\,\,$ & $\langle \Psi^{4/11}_{\rm CFD}\vert \Psi^{4/11}_{\rm CF-WYQ}\rangle$ & $\langle \Psi^{4/11}_{\rm CFD}\vert \Psi^{4/11}_{\rm L-R}\rangle$ & $\langle \Psi^{4/11}_{\rm CFD}\vert \Psi^{4/11}_{\rm L-MR}\rangle$ \\ \hline
 		12 & 5 & 1.0 & 1.0 & -- \\ 
 		16 &  6 & 0.9985(1) & 0.893(1) & 0.9977(0) \\ 
 		20 & 7 & 0.9834(1) & 0.976(1) & 0.9800(0) \\
 		24 & 8 & 0.9351(2) & 0.892(1) & 0.9551(4) \\
 		28 & 9 & 0.9627(2) & 0.790(3) & 0.9093(6) \\ \hline
 	\end{tabular}
 	\label{Table2}
 \end{table}

 \section{Phase Diagram}
 

 \begin{figure}[t]
 	\vspace{0.5cm}
 	\centering
 	\includegraphics[scale=0.6]{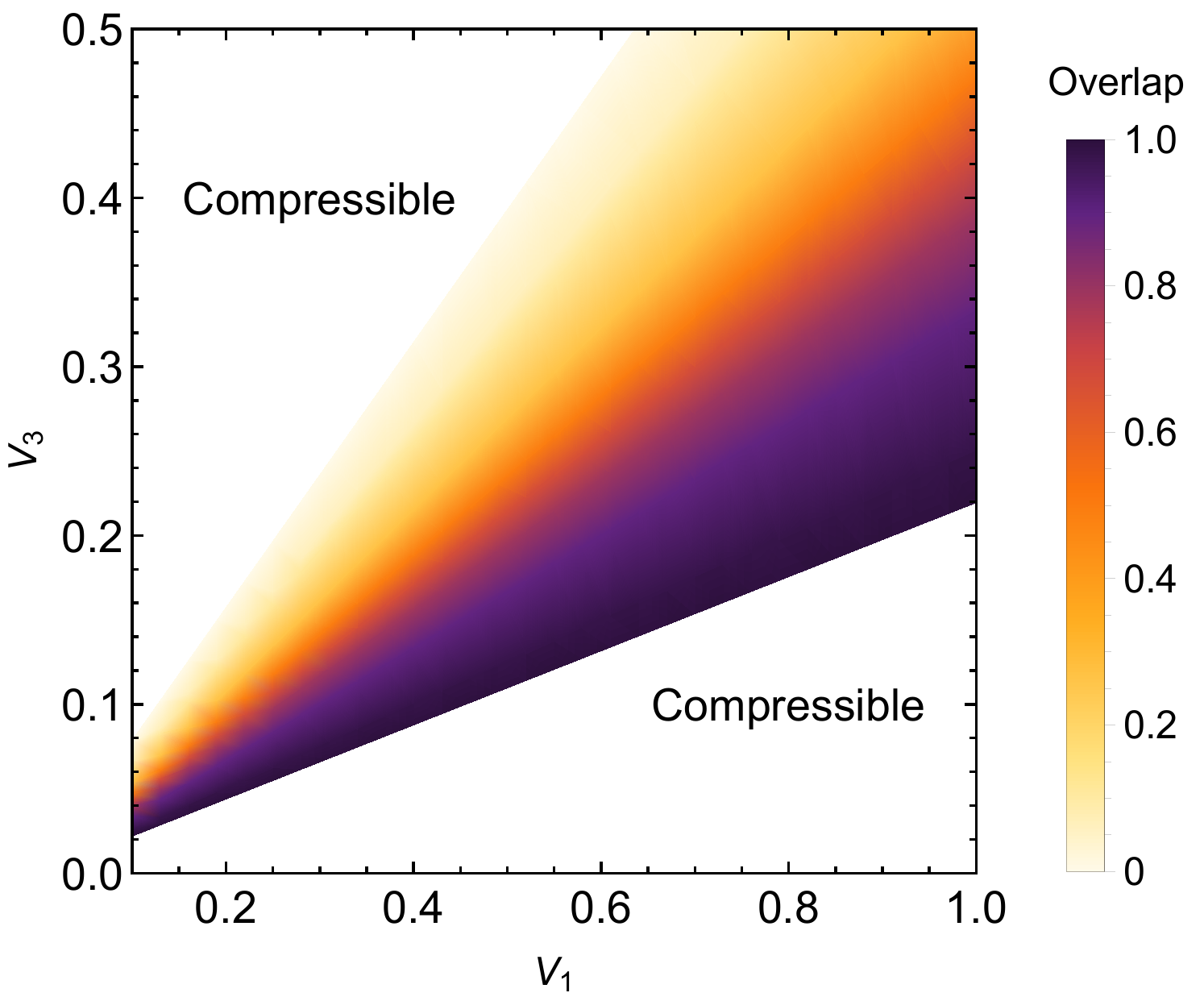}
 	\caption{(Color online) Phase Diagram for 4/11 state obtained for $N=12$ electrons. The filled region indicates the regime of incompressible ground state. The gradient in the color coding indicates the overlap of the exact ground state wave function in the hybrid pseudo-potentials of $V_1$ and $V_3$ with the CFD ground state wave function. We consider the phase as incompressible when the ground state is obtained at zero angular momentum and no other state is degenerate to this state. }
 	\label{Fig5}
 \end{figure}

We recall that the neighboring conventional states of $4/11$, i.e., $1/3$ and $2/5$ are understood through the model pseudo-potential $V_1$ only. 
On the other hand, the unconventional incompressible $4/11$ state is understood through a wave function which is partly constructed with a wave function that describes an incompressible state for $V_3$ pseudo-potential.
This indicates $4/11$ state should not be incompressible for $V_1$ alone, and $V_3$ pseudo-potential must have substantial role for the state's incompressibility. For  investigating whether or not this assertion is true, we obtain a phase diagram (Fig.~\ref{Fig5}) in  $V_1$--$V_3$ parameter space by performing exact diagonalization for 4/11 state with $N=12$ electrons and hybrid pseudo-potentials, and examining the state's nature. Clearly, unlike its neighboring conventional states, 4/11 state is not incompressible for $V_1$ alone. Also, $V_3$ alone cannot make the state $4/11$ incompressible. A window of $V_3/V_1 <1$
makes 4/11 state incompressible. 
Because the values of $V_1$ and $V_3$ for Coulomb potential in the lowest LL are in the same order of magnitude with $V_1 > V_3$, the unconventional states like $4/11$ and $5/13$ are incompressible along with their immediate neighboring conventional states $1/3$ and $2/5$.

\section{Discussion}  

In this paper, we have proposed a trial wave function for WYQ 1/3 state in terms of a product of Laughlin wave function and a ring function in which all electrons are correlated with two other electrons only. The entanglement spectra of this wave function 
as well as the exact wave function indicates that this state consists of two edge modes rather than one. This wave function has further been used for the 1/3 filled second $\Lambda$ level along with the completely filled the lowest $\Lambda$ level for constructing an wave function for 4/11 state. This indicates that $4/11$ state should have {\em three} gapless edge modes. It will be interesting to study the currents flowing through each of the channels and thereby determining the charge of the quasiparticle for 4/11 state. All of our results are based on the calculations on a sphere. We have also proposed analogous trial wave functions for a disk geometry. 

Another approach would be development of the conformal field theory for WYQ 1/3 state and thereby determining the ground-state wave function followed by proving its adiabatic connection with our proposed wave function.



\appendix

\section{Decomposition of a trial manybody wave function in single particle eigen basis}

Consider a general manybody wave function with total angular momentum $L=0$ in a spherical geometry as
\begin{equation}
\Psi = {\cal A}\prod_{i<j}^N (u_iv_j-u_jv_i)^{n_{ij}}
\label{Basis1}
\end{equation}
where $u_j = \cos(\theta_j /2)e^{i\phi_j /2}$ and $v_j = \sin(\theta_j /2)e^{i\phi_j /2}$ are the spherical spinors for $j^{\rm th}$ electron in terms of spherical angles $0\leq \theta_j \leq \pi$ and $0 \leq \phi_j \leq 2\pi$, the integer exponent $n_{ij} \geq 1$ (requirement due to Pauli exclusion principle) depends on pair of electrons with the constraint $\sum_j n_{ij} = N_\Phi$ for any electron with $N_\Phi$ being the total number of flux quanta. In general, $n_{ij}$ may not be equal for all $N(N-1)/2$ pairs and hence anti-symmetrization represented by ${\cal A}$ must be performed with different permutation of the pairs. We note that the ground state wave function of any FQHE state can be obtained by diagonalizing the manybody Hamiltonian in the linearly independent manybody basis functions (\ref{Basis1}) which may be obtained for different sets \cite{DDM} of $n_{ij}$ with the above constraint. Considering $\eta_j = u_j/v_j$, the wave function in (\ref{Basis1}) can be recast as
\begin{equation}
\Psi = \prod_j v_j^{N_\Phi} \Phi(\{\eta_j\})\,\, ; \,\, \Phi(\{\eta_j\})= {\cal A} \prod_{i<j}^N (\eta_i-\eta_j)^{n_{ij}} \, .
\label{Basis2}
\end{equation}
where $\Phi(\{\eta_j\})$ is an antisymmetric polynomial of degree $NN_\Phi /2$.
We then factor out Vandermonde determinant $V= \prod_{i<j}^N (\eta_i-\eta_j)$ from the function $\Phi(\{\eta_j\})$, i.e., 
\begin{equation}
\Phi(\{\eta_j\}) = V\tilde{\Phi}(\{\eta_j\}) \,\, ;\,\,  \tilde{\Phi}(\{\eta_j\}) = {\cal S} \prod_{i<j}^N (\eta_i-\eta_j)^{n_{ij}-1}
\label{Basis3}
\end{equation}
will be a symmetric polynomial of degree $N(N_\Phi-N+1)/2$, where ${\cal S}$ represents the symmetrization over the permutation of particles. 

The symmetric polynomial $\tilde{\Phi}(\{\eta_j\})$ can be expressed in  symmetric monomial \cite{Macdonald1979} basis $M_{\{\mu\}}$ given by
\begin{equation}
M_{\{\mu\}} = \sum_{P_{\{\mu \}}}\prod_{i=1}^N \eta_i^{\{\mu\}}
\label{Monomial}
\end{equation} 
with distinct set $\{\mu\} \equiv (\mu_1,\mu_2,\cdots,\mu_{_N})$ in descending order such that $0\leq \mu_i \leq N_\Phi-N+1$ and $\sum_{i=1}^N\mu_i = N(N_\Phi-N+1)/2$. Here the summation over $P_{\{\mu\}}$ represents all the permutations of the entries in the set $\{\mu \}$. We thus have \cite{MandalRay2018}
\begin{equation}
\tilde{\Phi}(\{\eta_j\}) = \sum_{\alpha} m_\alpha M_{\alpha\equiv \{ \mu \}}
\label{Basis4}
\end{equation}
where we associate a number $\alpha$ for a distinct set of $\{\mu\}$, the dimension of $\alpha$ is equal to the number of distinct symmetric polynomials,  and the coefficient $m_\alpha$ is the weight factor of the symmetric polynomial $M_\alpha$.

The symmetrization in Eq.(\ref{Basis3}) involves addition of $N!$ terms of the permutation, in general. It is a rather daunting task even with the use of Mathematica \cite{Mathematica} for algebraic manipulation with such a huge number of terms because each term consists of several monomial basis functions. We, however, exploit the form of the function $\prod_{i<j}^N (\eta_i-\eta_j)^{n_{ij}-1}$ for one of the terms of permutation to obtain $m_\alpha$ without explicit consideration of other $(N!-1)$ terms. 
We first expand $\prod_{i<j}^N (\eta_i-\eta_j)^{n_{ij}-1}$ as a sum of terms, each as products of the monomials in all the coordinates $\eta_i$. These terms belong to several groups identified with the appropriate set $\alpha$. By adding the coefficients of all the terms in a set $\alpha$, we find $m'_\alpha$.
We then inspect  entries of $\mu_i$ in the set $\alpha \equiv (\mu_1,\mu_2,\cdots,\mu_{_N})$ and count the number of equal entries. If all the $\mu_i$'s are different, $m'_\alpha$ gets renormalized for all the $N!$ terms. However, if there are $n_{\lambda}$ identical entries with $\lambda = 1,\cdots,k$, then $m'_\alpha$ normalizes to \begin{equation}
m_\alpha= m'_{\alpha}\prod_{\lambda =1}^{k} n_\lambda ! \, .
\end{equation}

The symmetric polynomial $\tilde{\Phi}(\{\eta_j\})$ can also be expanded in the Schur basis \cite{Macdonald1979} where the Schur functions are given by 
\begin{equation} \label{Schur}
S_{\{\mu\}}(\eta_1,\eta_2,\cdots,\eta_N) \equiv \frac{\begin{vmatrix}
	\eta_1^{\mu_1+N-1} & \eta_1^{\mu_2+N-2} & \cdots & \eta_1^{\mu_N} \\
	\eta_2^{\mu_1+N-1} & \eta_2^{\mu_2+N-2} & \cdots & \eta_2^{\mu_N} \\
	\vdots & \vdots & \ddots & \vdots \\
	\eta_N^{\mu_1+N-1} & \eta_N^{\mu_2+N-2} & \cdots & \eta_N^{\mu_N} 
	\end{vmatrix}}{\begin{vmatrix}
	\eta_1^{N-1} & \eta_1^{N-2} & \cdots & 1 \\
	\eta_2^{N-1} & \eta_2^{N-2} & \cdots & 1 \\
	\vdots & \vdots & \ddots & \vdots \\
	\eta_N^{N-1} & \eta_N^{N-2} & \cdots & 1
	\end{vmatrix}} \, .
\end{equation}
We thus have \cite{MandalRay2018}
\begin{equation}
\tilde{\Phi}(\{\eta_j\}) = \sum_{\alpha} s_{\alpha} S_{\alpha\equiv\{\mu\}} (\{\eta_j\}) \, ,
\label{Basis5}
\end{equation}
where the Schur coefficients $s_{\alpha}$ are to be determined.

From combinatorial theory, the Schur basis functions  (\ref{Schur}) is identified as a sum of monomials over semi-standard Young tableaux (SSYT) of shape $\alpha \equiv (\mu_1,\mu_2,\cdots,\mu_{_N})$ and which in turn related to monomial symmetric basis functions (\ref{Monomial}):
\begin{equation}\label{schur2mono}
S_\alpha (\eta_1,\eta_2,\cdots,\eta_{_N}) =  \sum_\beta K_{\alpha\beta} M_\beta (\eta_1,\eta_2,\cdots,\eta_{_N}) \, .
\end{equation}
The non-negative integer element $K_{\alpha\beta}$ is the number of SSYT of shape $\alpha$ and weight $\beta$ is called Kostka number\cite{Macdonald1979}. The matrix $K$ which we evaluate here using SageMath \cite{Sagemath} is an upper-triangular matrix with unity diagonal elements. Using Eqs.(\ref{Basis4}), (\ref{Basis5}) and (\ref{schur2mono}), we find
\begin{equation}
m_\beta = \sum_{\alpha} s_{\alpha} K_{\alpha\beta}
\end{equation}
which determines $s_\alpha$ as
\begin{equation}
s_\alpha = m_\alpha - \sum_{\beta =1}^{\alpha -1} s_\beta K_{\beta \alpha} \, .
\label{s_alpha}
\end{equation}

In the spherical geometry, the single particle eigenfunctions for the lowest Landau level are given by \cite{jain_book}
\begin{equation}
\phi_\lambda(u,v)= \left[ \frac{2Q+1}{4\pi} {2Q \choose Q-\lambda} \right]^{1/2} (-1)^{Q-\lambda} v^{Q-\lambda} u^{Q+\lambda}
\label{EqA1}
\end{equation}
where $ 2Q = N_\Phi $ is the total number of flux quanta, $\lambda =-Q,-Q+1,\cdots ,+Q$ indicate quantum numbers  of the degenerate states in the lowest Landau level. By defining $Q+\lambda =l$ and $\eta= u/v$, we find
\begin{equation}
\phi_l(u,v) = v^{2Q}\,{\cal N}_l \eta^l \, \, ;\,\, {\cal N}_l = \left[ \frac{2Q+1}{4\pi} \left( \begin{array}{c} 2Q \\ l  \end{array} \right) \right]^{1/2} (-1)^l 
\label{Single}
\end{equation}
with $l=0,1,\cdots,2Q$. Therefore, the Schur functions (\ref{Schur}) can be constructed with the single particle basis functions (\ref{Single}). The determinant in the numerator of the Schur functions are then related with the many electron basis functions which can be expressed in terms of single particle basis functions.

Equations (\ref{Basis2}), (\ref{Basis3}), (\ref{Schur}), (\ref{Basis5}), and (\ref{Single}) determine $\Psi$ with manybody determinant basis functions as
\begin{eqnarray}
\hspace{-10pt} \Psi &=& \sum_{\alpha}\frac{s_\alpha}{\prod_{i=1}^N {\cal N}_{\mu_i +N-i}}  \nonumber \\
&\times& \begin{vmatrix}
\phi_{\mu_1+N-1}(1) & \phi_{\mu_2+N-2}(1) & \cdots & \phi_{\mu_N}(1) \\
\phi_{\mu_1+N-1}(2) & \phi_{\mu_2+N-2}(2) & \cdots & \phi_{\mu_N}(2) \\
\vdots & \vdots & \ddots & \vdots \\
\phi_{\mu_1+N-1}(N) & \phi_{\mu_2+N-2}(N) & \cdots & \phi_{\mu_N} (N)
\end{vmatrix}
\label{psi_sp}
\end{eqnarray}
where $l=0,\cdots,2Q$ in $\phi_l(i)$ are the spherical quantum numbers, $(i)$ is the shorthand of $(u_i,v_i)$, and $\alpha \equiv (\mu_1,\cdots,\mu_{_N})$. \\

\noindent {\bf Example:} As an illustration for the use of the above algorithm, we consider Moore-Read wave function\cite{MR} in spherical geometry for $N=4$ with $N_\Phi = 2Q = 5$:
\begin{eqnarray}
\Psi_{\rm MR} &=& \sum_{i<j}^4 (u_iv_j-u_jv_i)^2 \, {\rm Pf}\left( \frac{1}{u_iv_j-u_jv_i}\right) \\
&=& \prod_{j=1}^4 v_j^5 \prod_{i<j}^4 \eta_{ij}^2 {\cal A} \left(\frac{1}{\eta_{13}\eta_{24}}  \right) \, ,
\end{eqnarray}
where $\eta_{ij} = \eta_i-\eta_j$ and ${\rm Pf}(A)$ represents Pfaffian of the antisymmetric matrix $A$. 
We thus find
\begin{equation}
\tilde{\Phi}(\{ \eta_j\}) = {\cal S} \left[\eta_{12}\eta_{14}\eta_{23}\eta_{34} \right] \, .
\end{equation}
As prescribed above, without performing explicit symmetrization here, we just consider the bracketed term above, i.e., the polynomial $P(\{\eta_j\}) = \eta_{12}\eta_{14}\eta_{23}\eta_{34}$. An expansion of $P(\{\eta_j\})$ yields
\begin{eqnarray}
\!\!\!\!\!\! P(\{\eta_j\}) &=& (-\eta_1^2\eta_3^2-\eta_2^2\eta_4^2) + (\eta_1^2\eta_2\eta_3-\eta_1^2\eta_2\eta_4 \nonumber \\
&& +\eta_1^2\eta_3\eta_4+\eta_4^2\eta_1\eta_2+\eta_4^2\eta_2\eta_3-\eta_4^2\eta_1\eta_3 \nonumber \\
&& -\eta_2^2\eta_1\eta_3+\eta_2^2\eta_1\eta_4 +\eta_2^2\eta_3\eta_4+\eta_3^2\eta_1\eta_4 \nonumber \\
&& +\eta_3^2\eta_1\eta_2-\eta_3^2\eta_2\eta_4 ) + (-2\eta_1\eta_2z\eta_3\eta_4) 
\label{Expansion}
\end{eqnarray}
which corresponds to three sets of $\{\mu\}$, namely, $(2,2,0,0)$, $(2,1,1,0)$, and $(1,1,1,1)$ denoted by $\alpha =1$, 2,and $3$ respectively. Therefore, $\tilde{\Phi}(\{ \eta_j\}) $ is the addition of $M_1$, $M_2$, and $M_3$ which respectively are the symmetric monomials constructed upon symmetrization of the respective group of terms within the parentheses in Eq.(\ref{Expansion}). We find $m'_1 = -2$, $m'_2=4$, and $m'_3 = -2$ which are the sum of the coefficients of these respective groups. In the sets, two entries occur twice, one entry occurs twice and two entries occur once, and one entry occurs four times respectively for $\alpha=1$, 2, and 3. Therefore, $m_1 = m'_1 \times 2!2!=-8$, $m_2 = m'_2 \times 2!1!1! = 8$, and $m_3 = m'_3 \times 4! = -48$.

We now determine the upper triangular $K$ matrix whose nonzero elements  $K_{\alpha\beta}$ are the Kostka numbers\cite{Macdonald1979} describing the number of SSYT possible of shape $\alpha$ and weight $\beta$, where  $\alpha$ and $\beta$ describe 3 sets, namely, $(2,2,0,0)$, $(2,1,1,0)$, and $(1,1,1,1)$. We thus find
\begin{equation}
K= \left(
\begin{array}{ccc}
1 & 1 & 2\\
0 & 1 & 3\\
0 & 0 & 1
\end{array} \right) \,.
\label{Koska}
\end{equation}
We next evaluate $s_\alpha$ using Eqs.~(\ref{s_alpha}) and (\ref{Koska}) and find $s_1 = -8$, $s_2 = 16$, and $s_3 =-80$. Using Eq.(\ref{psi_sp}), we therefore find (ignoring overall constant factor)
\begin{eqnarray}
\Psi_{MR} &\equiv& \frac{1}{\sqrt{3}} \begin{vmatrix}
\phi_{5}(1) & \phi_{4}(1) & \phi_1(1) & \phi_{0}(1) \\
\phi_{5}(2) & \phi_{4}(2) & \phi_1(2) & \phi_{0}(2) \\
\phi_{5}(3) & \phi_{4}(3) & \phi_1(3) & \phi_{0}(3) \\
\phi_{5}(4) & \phi_{4}(4) & \phi_1(4) & \phi_{0}(4)
\end{vmatrix} \nonumber \\
&&-\frac{1}{\sqrt{3}} \begin{vmatrix}
\phi_{5}(1) & \phi_{3}(1) & \phi_2(1) & \phi_{0}(1) \\
\phi_{5}(2) & \phi_{3}(2) & \phi_2(2) & \phi_{0}(2) \\
\phi_{5}(3) & \phi_{3}(3) & \phi_2(3) & \phi_{0}(3) \\
\phi_{5}(4) & \phi_{3}(4) & \phi_2(4) & \phi_{0}(4)
\end{vmatrix} \nonumber \\
&&+\frac{1}{\sqrt{3}} \begin{vmatrix}
\phi_{4}(1) & \phi_{3}(1) & \phi_2(1) & \phi_{1}(1) \\
\phi_{4}(2) & \phi_{3}(2) & \phi_2(2) & \phi_{1}(2) \\
\phi_{4}(3) & \phi_{3}(3) & \phi_2(3) & \phi_{1}(3) \\
\phi_{4}(4) & \phi_{3}(4) & \phi_2(4) & \phi_{1}(4)
\end{vmatrix}
\end{eqnarray}
which is precisely the same as the exact ground state description of three-body pseudo-potential for which Moore-Read \cite{MR} wave function is exact. \\

\section{Single Particle Basis Functions for First and Second $\Lambda$ levels in Spherical Geometry}

For the $^2$CFs, the effective monopole flux $2q = 2Q-2(N-1)$ and thus quantum numbers $m=-q,\,-q+1,\,\cdots,\, q$ and $m=-(q+1),\,-q,\,\cdots,\, (q+1)$ are present for single particle basis functions respectively in $\Lambda =0$ and $1$. 

The eigen basis function in $\Lambda =0$ are given by
\begin{equation}
\tilde{\phi}_m^{(0)}(u,v)= \left[ \frac{2q+1		}{4\pi} {2q \choose q-m} \right]^{1/2} (-1)^{q-m} v^{q-m} u^{q+m}
\end{equation}
which may be written for $j$-th particle as
\begin{equation}
\tilde{\phi}_l^{(0)} (j) \sim (-1)^l \left( \frac{1}{l!(2q-l)!}\right)^{1/2} v_j^{2q}\, \eta_j^l  
\label{Basis_lambda0}
\end{equation}
(up to $l$ dependent part of normalization factor)with $l=0,1,\cdots, 2q$ and $\eta_j=u_j/v_j$, by defining $l=q+m$.

The single particle eigen function in $\Lambda =1$ are given by
\begin{eqnarray}
\tilde{\phi}_m^{(1)}(u,v)&=& \left[ \frac{2q+3}{4\pi} \frac{(q+1-m)!(q+1+m)!}{(2q+1)!} \right]^{1/2} \nonumber \\
 &\times& \left[ {2q+1 \choose q+1-m}vv^\ast - {2q+1 \choose q+1+m}uu^\ast \right] \nonumber \\
&\times& (-1)^{q+1-m} v^{q-m} u^{q+m}
\end{eqnarray}
with $m=-(q+1),\, -q,\,\cdots,\, (q+1)$; while the first term in the bracket is excluded for $m=-(q+1)$, the second term is excluded for $m=q+1$.
The lowest-Landau-level projected basis functions then found to be for $j$-th particle \cite{Mandal2018} as 
\begin{eqnarray}
\hspace{-17pt} \tilde{\phi}_{l,\rm proj}^{(1)}&=& {\cal N}_l^{(1)} v_j^{2q} \eta_j^l \nonumber \\
& \times & \left[ {2q+1 \choose l-1}\left(Q_j - \frac{N-1}{\eta_j} \right) + {2q+1 \choose l}Q_j \right]
\end{eqnarray}
with $l=q+1+m$, i.e., $l=0,1,\cdots ,\,2(q+1)$, ($l\neq 0$ for the first term  and $l\neq 2(q+1)$ in the second term inside the bracket), $\eta_j = u_j/v_j$, $Q_j = \sum_{k\neq j} (v_jv_k)/(u_jv_k-u_kv_j)$, and ${\cal N}_l^{(1)}= \left[ \frac{2q+3}{4\pi}\frac{(2q+2-l)!l!}{(2q+1)!}  \right]^{1/2} (-1)^{l}$. Now dropping the functions which are already included in the basis functions of $\Lambda =0$, we find the basis states (keeping only the $l$ dependent constant factor of the normalization constant),
\begin{eqnarray}
\tilde{\phi}_{l,\rm proj}^{(1)} (j)&\sim& (-1)^l \left(\frac{1}{l!(2q+2-l)!}\right)^{1/2} v_j^{2q} \eta_j^l \nonumber \\
& \times&  \left\{  \begin{array}{ll} l(
Q_j- \frac{N-1}{\eta_j}) & {\rm for} \,\,\,l= 2(q+1) \\
Q_j  & {\rm otherwise}
\end{array}\right.
\label{Basis_lambda1}
\end{eqnarray}
We note that Eqs.(\ref{Basis_lambda0}) and (\ref{Basis_lambda1}) have close resemblance with the corresponding wave functions in disc geometry.\cite{jain_book}

\section{Conversion of Many body wave function for lowest Landau level to $\Lambda=1$ level }

Using the method of DSPEB developed above, we can decompose $\Psi^{1/3,N^\ast}_{\rm L-R}$ for $N^\ast$ particles in terms of single particle basis $\tilde{\phi}_l^{(0)}$ (see Eq.(\ref{Basis_lambda0}))  of $\Lambda=0$ as
\begin{equation}
\left. \Psi^{1/3,N^\ast}_{\rm L-R}\right\vert_{\Lambda=0} = \sum_{\alpha} C_\alpha 
 \begin{vmatrix}
\tilde{\phi}^{(0)}_{l_1}(1) & \tilde{\phi}^{(0)}_{l_2}(1) & \cdots & \tilde{\phi}^{(0)}_{l_{N^\ast}}(1) \\
\tilde{\phi}^{(0)}_{l_1}(2) & \tilde{\phi}^{(0)}_{l_2}(2) & \cdots & \tilde{\phi}^{(0)}_{l_{N^\ast}}(2) \\
\vdots &  \vdots & \vdots & \vdots \\
\tilde{\phi}^{(0)}_{l_1}(N^\ast) & \tilde{\phi}^{(0)}_{l_2}(N^\ast) & \cdots & \tilde{\phi}^{(0)}_{l_{N^\ast}}(N^\ast) 
\end{vmatrix}_\alpha
\label{Ring_lambda0}
\end{equation} 
For constructing $\Psi^{1/3,N^\ast}_{\rm L-R}$-like wave function in $\Lambda =1$, we need to replace $\tilde{\phi}_{l_i}^{(0)}$ by $\tilde{\phi}_{l_i}^{(1)}$ in Eq.(\ref{Ring_lambda0}). Now exploiting the common terms in the expression of $\tilde{\phi}_{l_i}^{(0)}$ and $\tilde{\phi}_{l_i}^{(1)}$ (see Eqs.(\ref{Basis_lambda0}) and (\ref{Basis_lambda1})), we
are able to express 
\begin{eqnarray}
\left. \!\!\!\!\!\!\! \Psi^{1/3,N^\ast}_{\rm L-R}\right\vert_{\Lambda=1} &=& \prod_{j=1}^{N^\ast} Q_j \sum_{\alpha} D_\alpha \nonumber \\
&\times& \begin{vmatrix}
\tilde{\phi}^{(0)}_{l_1}(1) & \tilde{\phi}^{(0)}_{l_2}(1) & \cdots & \tilde{\phi}^{(0)}_{l_{N^\ast}}(1) \\
\tilde{\phi}^{(0)}_{l_1}(2) & \tilde{\phi}^{(0)}_{l_2}(2) & \cdots & \tilde{\phi}^{(0)}_{l_{N^\ast}}(2) \\
\vdots &  \vdots & \vdots & \vdots \\
\tilde{\phi}^{(0)}_{l_1}(N^\ast) & \tilde{\phi}^{(0)}_{l_2}(N^\ast) & \cdots & \tilde{\phi}^{(0)}_{l_{N^\ast}}(N^\ast)
\end{vmatrix}_\alpha
\label{Ring_lambda1}
\end{eqnarray}
with an exception that $\tilde{\phi}^{(0)}_{2q+2}(j)$ should be multiplied by the factor $1 - (N-1)/(\eta_jQ_j)$ 
\begin{equation}
D_\alpha = C_\alpha \prod_{i=1}^{N^\ast} \left( \frac{1}{(2q+2-l_i)(2q+1-l_i)}\right)^{1/2} \ .
\end{equation}

\section*{Acknowledgments} We thank Ajith Balram for pointing out a mistake in an early version of the manuscript which was circulated before its first submission for publication. We thank Sutirtha Mukherjee for generously sharing us the published CFD data. We acknowledge the
Param Shakti (IIT Kharagpur)--a National Supercomputing Mission, Government of India for providing their computational resources. 
SSM is supported by  the Council of Scientific and Industrial Research, Human Resource Development Group, India, through its Scheme No.: 03(1436)/18/EMR-II. \\


\end{document}